\documentclass{article}
\usepackage{arxiv}
\usepackage{tikz}
\usepackage{algorithm}
\usepackage{algpseudocode}
\usetikzlibrary{
  shapes,
  arrows,
  arrows.meta,
  positioning,
  calc,
  backgrounds,
  fit,
  decorations.pathreplacing,
  patterns
}
\usepackage{xcolor}
\definecolor{proteinblue}{RGB}{66, 133, 244}
\definecolor{ligandorange}{RGB}{251, 188, 5}
\definecolor{complexgreen}{RGB}{52, 168, 83}
\definecolor{fusionpurple}{RGB}{156, 39, 176}
\definecolor{outputgray}{RGB}{95, 99, 104}
\definecolor{affinityteal}{RGB}{0, 150, 136}
\definecolor{kineticspink}{RGB}{233, 30, 99}
\definecolor{sagold}{RGB}{255, 193, 7}
\definecolor{gcdmblue}{RGB}{33, 150, 243}
\usepackage[utf8]{inputenc}    
\usepackage[T1]{fontenc}       
\usepackage{amsmath,amssymb,amsfonts}   
\usepackage{graphicx}
\usepackage{booktabs}
\usepackage{microtype}         
\usepackage{nicefrac}         
\usepackage{doi}
\usepackage{hyperref}          
\usepackage{newunicodechar}
\newunicodechar{∼}{\textasciitilde}
\usepackage{dsfont}            
\usepackage{enumitem}          
\usepackage{subcaption}        
\usepackage{float}             
\usepackage{grffile}           
\usepackage{url}               
\usepackage{natbib}            
\usepackage{setspace}

\title{KinetiDiff: Docking-Guided Diffusion for De Novo ACVR1 Inhibitor Design in Fibrodysplasia Ossificans Progressiva}

\author{Aaryan Patel \\
Saugus High School, Santa Clarita, CA \\
\texttt{aaryanp0302@gmail.com}
}
\begin{document}
\pagestyle{fancy}
\fancyhead{}                    
\renewcommand{\headrulewidth}{0pt} 
\maketitle

\begin{abstract}
We present \textbf{KinetiDiff}, a structure-based framework for de novo kinase inhibitor design that integrates a Geometry-Complete Diffusion Model with real-time AutoDock Vina gradient guidance. By injecting physics-based docking gradients into the diffusion denoising loop, KinetiDiff steers molecule generation toward high-affinity conformations for ACVR1 (ALK2), the causative kinase in Fibrodysplasia Ossificans Progressiva.

From 10,000 diffusion samples, the framework produced 9,997 valid molecules. The best candidate achieved $-11.05$~kcal/mol (pKd = 8.10), a 19.2\% improvement over the crystallographic reference. The top 100 candidates all exceed the reference, with 100\% Lipinski compliance, median synthetic accessibility of 2.67, and internal diversity of 0.790.

Systematic ablation across four guidance strategies—Vina-Direct (physics), HNN-Denovo (neural proxy), multi-objective, and unguided—demonstrates that real-time docking guidance dominates on all metrics. We evaluate HNN-Denovo as a computationally efficient alternative (60-fold speedup per step), revealing a domain-mismatch limitation ($r = 0.224$ correlation with Vina) that explains its inferior performance.

These results establish gradient-guided geometric diffusion as a practical approach for generating potent, synthetically accessible inhibitors against rare-disease kinase targets.
\end{abstract}

\section{Introduction}

\textbf{Fibrodysplasia Ossificans Progressiva (FOP)} is a rare autosomal-dominant disorder caused by gain-of-function mutations in \textit{ACVR1}, the gene encoding the ALK2 kinase. The most prevalent mutation, R206H, renders ALK2 constitutively active and responsive to activin~A, triggering aberrant osteogenic signaling through SMAD1/5/8 phosphorylation \citep{Kaplan2008FOP,Srinivasan2024Activin}. This hyperactivation drives progressive heterotopic ossification of soft tissues—muscles, tendons, ligaments—resulting in cumulative immobility and severe morbidity.


Current ALK2 inhibitors—saracatinib (repurposed Src inhibitor), zilurgisertib (selective ALK2 inhibitor in trials), and palovarotene (indirect mechanism)—have shown efficacy but carry limitations: off-target activity, unresolved safety profiles, and incomplete mechanistic targeting \citep{Williams2021SaracatinibFOP,Ullrich2025Zilurgisertib}. These candidates emerged from traditional medicinal chemistry, exploring only a narrow slice of tractable chemical space.

\textit{Generative models} offer an alternative: rather than screening libraries, they design molecules de novo by learning drug-like distributions and sampling conditioned on target constraints \citep{Zhavoronkov2019DDR1,Li2022RIPK1,Das2025GenerativeReview}. \textit{Diffusion models} are particularly effective for structure-conditioned ligand generation because they operate in 3D coordinate space with geometric inductive biases that preserve molecular symmetries \citep{Morehead2024geometry,Zheng2024Apo2Mol,Zhang2025GCLDM,Jian2024BindingAffinityGuidance,Hu2025GuidedEquivariant,Weller2024DrugHIVE}.

However, most diffusion-based drug design pipelines decouple generation from binding evaluation: molecules are sampled unconditionally or with pocket-shape conditioning, then filtered post hoc using docking or property proxies. The diffusion model never sees the actual docking score during sampling, limiting its ability to steer toward high-affinity solutions. While recent work on classifier-guided diffusion has begun addressing this \citep{Jian2024BindingAffinityGuidance,Hu2025GuidedEquivariant}, no prior framework has integrated real-time AutoDock Vina scoring directly into denoising for a therapeutic target.

We present \textbf{KinetiDiff}, which injects AutoDock Vina docking gradients into the reverse diffusion process. By computing numerical gradients of the Vina scoring function at each denoising step, KinetiDiff guides molecule generation toward high-affinity conformations for the ACVR1 binding pocket. We evaluate KinetiDiff against three alternatives—HNN-Denovo (neural proxy, 60$\times$ speedup), multi-objective scoring, and unguided generation—demonstrating superior performance across all metrics. Applied to ACVR1, the framework generates 9,997 valid molecules, with the best achieving $-11.05$~kcal/mol—a 19.2\% improvement over the crystallographic reference.

\section{Background: Diffusion Models for Molecular Generation}
\label{sec:background}

This section presents the mathematical foundations of denoising diffusion probabilistic models (DDPMs) and their extension to 3D molecular generation through the Geometry-Complete Diffusion Model (GCDM).

\subsection{Forward Diffusion Process}

A diffusion model defines a Markov chain that gradually corrupts data $\mathbf{x}_0$ by adding Gaussian noise over $T$ timesteps. At each step, a small amount of noise is injected according to a variance schedule $\{\beta_t\}_{t=1}^T$:
\begin{equation}
q(\mathbf{x}_t \mid \mathbf{x}_{t-1}) = \mathcal{N}\!\left(\mathbf{x}_t;\, \sqrt{1-\beta_t}\,\mathbf{x}_{t-1},\, \beta_t\mathbf{I}\right).
\end{equation}

A key property of this Markov chain is that the marginal $q(\mathbf{x}_t \mid \mathbf{x}_0)$ can be written in closed form. Defining $\alpha_t = 1-\beta_t$ and $\bar{\alpha}_t = \prod_{s=1}^{t}\alpha_s$, repeated application of the reparameterization trick yields
\begin{equation}
q(\mathbf{x}_t \mid \mathbf{x}_0) = \mathcal{N}\!\left(\mathbf{x}_t;\, \sqrt{\bar{\alpha}_t}\,\mathbf{x}_0,\, (1-\bar{\alpha}_t)\mathbf{I}\right),
\label{eq:forward-marginal}
\end{equation}
which enables direct sampling of any noised version of $\mathbf{x}_0$ without iterating through intermediate steps:
\begin{equation}
\mathbf{x}_t = \sqrt{\bar{\alpha}_t}\,\mathbf{x}_0 + \sqrt{1-\bar{\alpha}_t}\,\boldsymbol{\epsilon}, \quad \boldsymbol{\epsilon} \sim \mathcal{N}(\mathbf{0},\mathbf{I}).
\label{eq:reparam}
\end{equation}

In KinetiDiff, we adopt a cosine noise schedule \citep{Kingma2021Improved} with offset $s = 0.008$:
\begin{equation}
\bar{\alpha}_t = \frac{\cos^2\!\left(\frac{t/T + s}{1+s}\cdot\frac{\pi}{2}\right)}{\cos^2\!\left(\frac{s}{1+s}\cdot\frac{\pi}{2}\right)},
\label{eq:cosine-schedule}
\end{equation}
which provides a smooth transition from data to noise and avoids the abrupt signal collapse observed with linear schedules, improving sample quality for structured outputs such as molecular coordinates \citep{Ho2020DDPM,Sohl2015Nonequilibrium}.

\subsection{Reverse Process and Score Matching}

Generation proceeds by learning to reverse the forward corruption. The reverse transition is parameterized as
\begin{equation}
p_\theta(\mathbf{x}_{t-1} \mid \mathbf{x}_t) = \mathcal{N}\!\left(\mathbf{x}_{t-1};\, \boldsymbol{\mu}_\theta(\mathbf{x}_t, t),\, \sigma_t^2\mathbf{I}\right),
\end{equation}
where $\sigma_t^2$ follows the fixed-variance schedule of \citet{Ho2020DDPM}. The learned mean takes the form
\begin{equation}
\boldsymbol{\mu}_\theta(\mathbf{x}_t, t) = \frac{1}{\sqrt{\alpha_t}}\left(\mathbf{x}_t - \frac{\beta_t}{\sqrt{1-\bar{\alpha}_t}}\,\boldsymbol{\epsilon}_\theta(\mathbf{x}_t, t)\right),
\label{eq:reverse-mean}
\end{equation}
where $\boldsymbol{\epsilon}_\theta$ is a neural network trained to predict the noise added at step $t$. Training minimizes the denoising score-matching objective \citep{Ho2020DDPM,Song2019ScoreMatching,Song2020ScoreSDE}:
\begin{equation}
\mathcal{L}_{\text{DSM}} = \mathbb{E}_{t \sim \mathcal{U}(1,T),\;\boldsymbol{\epsilon}\sim\mathcal{N}(\mathbf{0},\mathbf{I})}\!\left[\left\|\boldsymbol{\epsilon} - \boldsymbol{\epsilon}_\theta\!\left(\sqrt{\bar{\alpha}_t}\,\mathbf{x}_0 + \sqrt{1-\bar{\alpha}_t}\,\boldsymbol{\epsilon},\;t\right)\right\|^2\right].
\label{eq:dsm}
\end{equation}

The connection to score-based generative modeling is direct: the noise predictor $\boldsymbol{\epsilon}_\theta$ is related to the score function $\nabla_{\mathbf{x}_t}\log p(\mathbf{x}_t)$ by
\begin{equation}
\boldsymbol{\epsilon}_\theta(\mathbf{x}_t, t) \approx -\sqrt{1-\bar{\alpha}_t}\,\nabla_{\mathbf{x}_t}\log p(\mathbf{x}_t).
\label{eq:score-connection}
\end{equation}
This relationship is central to guidance: by modifying the score with an auxiliary gradient (Section~\ref{sec:vina-guidance}), we can steer generation toward regions of chemical space with desirable properties—in our case, high ACVR1 binding affinity.

\subsection{GCDM: Geometry-Complete Extensions for 3D Molecules}
\label{sec:gcdm}

Standard DDPMs operate on flat feature vectors. Molecules in 3D, however, possess geometric structure that must be respected: a rotation or translation of the entire molecule should not change the generated distribution. The Geometry-Complete Diffusion Model (GCDM) \citep{Morehead2024geometry} addresses this through three key modifications.

\paragraph{SE(3)-Equivariant Denoising.}
The noise predictor $\boldsymbol{\epsilon}_\theta$ is implemented as an SE(3)-equivariant graph neural network. Atomic coordinates $\mathbf{X}\in\mathbb{R}^{N\times 3}$ transform equivariantly ($\boldsymbol{\epsilon}_\theta(R\mathbf{X}+\mathbf{t}, \mathbf{h}, t) = R\,\boldsymbol{\epsilon}_\theta(\mathbf{X}, \mathbf{h}, t)$), while atom features $\mathbf{h}$ (element type, formal charge) remain invariant. This is realized through the Geometry-Complete Perceptron (GCP), which decomposes hidden representations into \textit{scalar} channels $\mathbf{s}\in\mathbb{R}^{d_s}$ (invariant) and \textit{vector} channels $\mathbf{v}\in\mathbb{R}^{3\times d_v}$ (equivariant). Message passing updates both channels while preserving symmetry:
\begin{align}
\mathbf{s}_i^{(\ell+1)} &= \mathbf{s}_i^{(\ell)} + \text{MLP}\!\left(\mathbf{s}_i^{(\ell)}, \|\mathbf{v}_i^{(\ell)}\|^2, \sum_{j \in \mathcal{N}(i)} \phi_s(\mathbf{s}_j^{(\ell)}, d_{ij})\right), \\
\mathbf{v}_i^{(\ell+1)} &= \mathbf{v}_i^{(\ell)} + \sum_{j \in \mathcal{N}(i)} \phi_v(\mathbf{s}_j^{(\ell)}, d_{ij}) \cdot \frac{\mathbf{x}_j - \mathbf{x}_i}{\|\mathbf{x}_j - \mathbf{x}_i\| + \varepsilon},
\end{align}
where:
\begin{itemize}[nosep,leftmargin=1.5em]
    \item $\mathbf{s}_i^{(\ell)}, \mathbf{v}_i^{(\ell)}$ -- scalar and vector hidden states of atom $i$ at layer $\ell$
    \item $\mathcal{N}(i)$ -- set of neighbors of atom $i$ within the graph cutoff radius ($10$~\AA)
    \item $d_{ij} = \|\mathbf{x}_i - \mathbf{x}_j\|$ -- interatomic Euclidean distance
    \item $\phi_s, \phi_v$ -- learnable scalar and vector edge message functions
    \item $\varepsilon$ -- small constant for numerical stability
\end{itemize}
The vector channel encodes directional information (e.g., bond orientations), while the scalar channel aggregates distance-based features. Because all operations depend only on relative positions and invariant norms, the overall architecture is SE(3)-equivariant by construction.

\paragraph{Pocket Conditioning.}
To generate molecules complementary to a specific binding site, GCDM conditions the denoising process on the 3D structure of the protein pocket. Given the ACVR1 crystal structure (PDB: 3MTF), we define the pocket neighborhood as:
\begin{equation}
\mathcal{N}_c = \left\{ (\mathbf{v}_i, \mathbf{r}_i) \;\middle|\; \| \mathbf{r}_i - \mathbf{r}_c \| \leq 10\,\text{\AA},\; \forall\, i \in \mathcal{I} \right\},
\label{eq:pocket-neighborhood}
\end{equation}
where:
\begin{itemize}[nosep,leftmargin=1.5em]
    \item $\mathbf{r}_c = (24.87, -12.54, 38.40)$~\AA{} -- binding site centroid
    \item $\mathbf{r}_i$ -- C$_\alpha$ coordinate of residue $i$
    \item $\mathbf{v}_i$ -- residue feature vector (amino acid type, secondary structure)
    \item $\mathcal{I}$ -- set of all residue indices in the receptor
\end{itemize}
Each residue $i$ receives a distance-dependent weight $w_i \propto \exp(-\gamma\|\mathbf{r}_i - \mathbf{r}_c\|^2)$, and the pocket embedding is formed by weighted pooling of SE(3)-equivariant residue features followed by concatenation with sinusoidal timestep embeddings:
\begin{equation}
\mathbf{c}_t = \text{MLP}([\textstyle\sum_{i\in\mathcal{P}} w_i\,\mathbf{h}_i^{(L)};\;\mathbf{e}(t)]).
\end{equation}
This conditioning vector is injected additively into each denoising layer, providing the model with spatial context about the target binding site while preserving SE(3) invariance of the overall generation process.

\paragraph{Relevance to FOP.}
The ACVR1 R206H mutation alters the GS-domain activation loop adjacent to the ATP-binding cleft. Pocket-conditioned generation enables GCDM to produce molecules complementary to this specific 3D geometry—essential for rare-disease targets with limited known scaffolds and high chemical exploration requirements.

\section{Methodology}
\label{sec:methods}

\subsection{Overview}
\begin{figure*}[t]
\centering
\includegraphics[width=\textwidth]{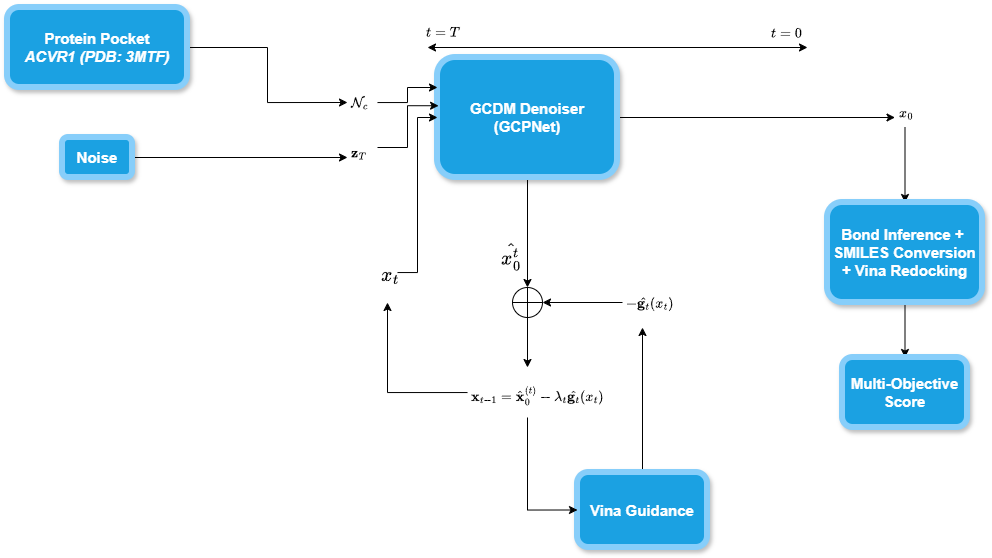}
\caption{Architecture overview of the KinetiDiff framework. \textbf{Left:} Inputs---protein pocket (ACVR1, PDB: 3MTF) and initial noise $\mathbf{z}_T \sim \mathcal{N}(\mathbf{0}, \mathbf{I})$. \textbf{Center:} Iterative reverse diffusion loop with the GCDM denoiser (SE(3)-equivariant GCPNet) and real-time Vina gradient guidance injected at each qualifying timestep via $\mathbf{x}_{t-1} = \hat{\mathbf{x}}_0^{(t)} - \lambda_t \hat{\mathbf{g}}_t$. \textbf{Right:} Post-processing pipeline---bond inference, SMILES canonicalization, Vina redocking, and multi-objective scoring. Created with BioRender \citep{BioRender2026}.}
\label{fig:architecture}
\end{figure*}

KinetiDiff consists of three stages: (i) pocket-conditioned 3D molecule generation via GCDM, (ii) real-time gradient guidance from AutoDock Vina during the reverse diffusion process, and (iii) post-hoc multi-objective ranking. The framework is implemented in PyTorch Lightning \citep{Falcon2019Lightning}; generation and guidance are unified in a single sampling loop (Figure~\ref{fig:architecture}).

Generation begins by sampling initial molecular representations from an isotropic Gaussian:
\begin{equation}
\mathbf{z}_T = (\mathbf{x}_T, \mathbf{h}_T) \sim \mathcal{N}(\mathbf{0}, \mathbf{I}), \quad \mathbf{x}_T \in \mathbb{R}^{N \times 3}, \quad \mathbf{h}_T \in \mathbb{R}^{N \times 9},
\label{eq:initial-noise}
\end{equation}
where:
\begin{itemize}[nosep,leftmargin=1.5em]
    \item $\mathbf{x}_T$ -- atomic coordinates ($N$ atoms in 3D space), sampled in the zero center-of-mass subspace
    \item $\mathbf{h}_T$ -- atom type features (9-dimensional one-hot encoding over \{C, N, O, F, P, S, Cl, Br, I\})
    \item $N$ -- number of atoms, sampled from the empirical size distribution of the training set
\end{itemize}
The coordinate samples are projected to the zero center-of-mass subspace, reducing the effective dimensionality to $(N{-}1) \times 3$ degrees of freedom. The reverse diffusion process then iteratively denoises $\mathbf{z}_T$ toward a valid molecular structure conditioned on the ACVR1 pocket (Eq.~\ref{eq:pocket-neighborhood}).

\subsection{Real-Time Vina Gradient Guidance}
\label{sec:vina-guidance}

The central innovation of KinetiDiff is the injection of AutoDock Vina \citep{Trott2010AutoDock} docking gradients directly into the denoising loop. At selected timesteps during reverse diffusion, we evaluate the current (partially denoised) molecular structure against the ACVR1 pocket and shift atomic coordinates in the direction that improves binding affinity:
\begin{equation}
\mathbf{x}_{t-1} = \hat{\mathbf{x}}_0^{(t)} - \lambda_t\,\hat{\mathbf{g}}_t,
\label{eq:guided-update}
\end{equation}
where:
\begin{itemize}[nosep,leftmargin=1.5em]
    \item $\hat{\mathbf{x}}_0^{(t)}$ -- the GCDM denoiser's predicted clean coordinates at timestep $t$ (standard denoising output before guidance)
    \item $\lambda_t$ -- timestep-dependent guidance strength (Eq.~\ref{eq:guidance-schedule})
    \item $\hat{\mathbf{g}}_t$ -- numerical Vina gradient, normalized and clipped (see Appendix~\ref{appendix:formulas})
\end{itemize}

\paragraph{Numerical Gradient Computation.}
Because Vina's scoring function is not natively differentiable, we compute gradients via central finite differences. The full gradient vector is assembled over all atoms and spatial dimensions:
\begin{equation}
\hat{\mathbf{g}}_t = \sum_{i=1}^{N}\sum_{d \in \{x,y,z\}} \left( \frac{\mathcal{U}_{\text{Vina}}(\mathbf{x}_t + \varepsilon\,\mathbf{e}_{i,d}) - \mathcal{U}_{\text{Vina}}(\mathbf{x}_t - \varepsilon\,\mathbf{e}_{i,d})}{2\varepsilon} \right) \mathbf{e}_{i,d},
\label{eq:finite-diff}
\end{equation}
where:
\begin{itemize}[nosep,leftmargin=1.5em]
    \item $\mathcal{U}_{\text{Vina}}(\cdot)$ -- AutoDock Vina scoring function (kcal/mol; more negative = stronger binding)
    \item $\varepsilon = 0.05$~\AA{} -- finite-difference step size
    \item $\mathbf{e}_{i,d}$ -- unit perturbation vector for atom $i$ along spatial dimension $d$
\end{itemize}
Each gradient evaluation requires $6N+1$ Vina scoring calls for an $N$-atom molecule. Gradients are clipped to a maximum norm of 10.0 and then normalized to unit length after scaling (Appendix~\ref{appendix:formulas}).

\paragraph{Timestep-Dependent Scaling.}
Guidance strength is modulated by an exponential schedule with clipping bounds and an activation indicator:
\begin{equation}
\lambda_t = \left[ 0.1 \cdot \exp\!\left(3\!\left(1 - \frac{t}{T}\right)\right) \right]_{0.1}^{10.0} \cdot \mathds{1}_{[t \geq 0.2T]},
\label{eq:guidance-schedule}
\end{equation}
where:
\begin{itemize}[nosep,leftmargin=1.5em]
    \item $[\cdot]_{a}^{b}$ denotes clamping to the interval $[a, b]$
    \item $\mathds{1}_{[t \geq 0.2T]}$ is an indicator function that activates guidance only during the last 80\% of the reverse process
    \item At $t=T$ (pure noise), $\lambda \approx 0.1$; at $t=0$ (final sample), $\lambda \approx 2.0$
\end{itemize}
This schedule suppresses guidance during early (noisy) timesteps where the molecule lacks coherent geometry and Vina scores would be meaningless. The indicator disables guidance for the final 20\% of generation, allowing the diffusion model to refine fine-grained chemical details (bond lengths, atom types) without interference. Within the active window, the exponential ramp concentrates guidance effort on the middle-to-late denoising phase where molecular structure is emerging but not yet finalized.

\paragraph{Docking Configuration.}
All guidance-phase Vina evaluations use local docking mode (exhaustiveness~1, single output mode) centered at the ACVR1 binding site centroid $(24.87, -12.54, 38.40)$~\AA{} with a $22\times22\times22$~\AA{} search box and fixed seed~42.

\subsection{HNN-Denovo: A Lightweight Alternative}
\label{sec:hnn}

As an alternative to the computationally expensive Vina calls, we also evaluate a learned guidance signal. HNN-Denovo is a convolutional neural network trained on ACVR1-family binding data from BindingDB \citep{Liu2007BindingDB} (15,324 protein--ligand pairs, of which 2,847 target ACVR1 directly). The network encodes protein sequences and ligand SMILES through parallel 1D convolutional branches (kernels $k \in \{3,5,7\}$, 256 channels each), fuses the representations, and predicts binding affinity (pKd).

To enable gradient-based guidance, HNN-Denovo computes a differentiable proxy affinity score from 3D coordinates using soft BINANA-inspired interaction descriptors: smooth contact counts via $\sigma((\tau - d_{ij})\cdot\kappa)$ (where $\tau = 4.5$~\AA{} and $\kappa = 2.0$), and inverse-distance proximity potentials. These terms are fully differentiable with respect to atomic positions, enabling direct backpropagation without finite-difference approximation. Guidance gradients are clipped to a maximum norm of 10.0 and scaled linearly with timestep fraction.

HNN-Denovo runs 60$\times$ faster per guidance step, as it requires only a forward pass and backpropagation rather than external docking calls. The model achieves strong test-set performance (PCC~$= 0.72$, RMSE~$= 0.70$ pKd), confirming accurate sequence--ligand representations. However, gradient guidance must rely on a proxy score computed from 3D atomic positions (centroid distance, contact counts, potentials; Appendix~\ref{sec:proxy}), which achieves only $r = 0.224$ correlation with Vina. The proxy mismatch between the model's SMILES-based input and the diffusion's coordinate space limits guidance accuracy; Vina-Direct, operating natively on 3D coordinates via finite differences, is essential for docking accuracy.

\subsection{Multi-Objective Scoring}
\label{sec:scoring}

Generated molecules are ranked using a composite score:
\begin{equation}
S(\mathbf{x}) = 0.5 \cdot S_{\text{aff}}(\text{pKd}) + 0.3 \cdot S_{\text{kin}}(k_{\text{off}}) + 0.2 \cdot S_{\text{SA}}(\text{SA}),
\end{equation}
where the affinity component is a Gaussian (pKd $= 7.5$, $\sigma = 1.5$), the kinetics component rewards $k_{\text{off}} \in [0.1, 1.0]$~s$^{-1}$ via box-exponential weighting, and the SA component penalizes beyond SA $= 4.0$. Dissociation rates are estimated via $\log_{10}(k_{\text{off}}) \approx -0.5\,\text{pKd} + 3.0$. Pareto-optimal molecules are identified to provide diverse trade-off solutions.

\subsection{Docking Validation Protocol}
\label{sec:docking-protocol}

All final candidate molecules were uniformly redocked against ACVR1 (PDB: 3MTF) using AutoDock Vina 1.2.3 under identical conditions: search box $22 \times 22 \times 22$~\AA{} centered at $(24.87, -12.54, 38.40)$~\AA, exhaustiveness~4, seed~42. 3D conformers were generated using RDKit's ETKDGv3 (with fallback to v2 and random coordinate embedding), energy-minimized with MMFF94, and converted to PDBQT format via Meeko. The crystallographic reference inhibitor from 3MTF was redocked as a baseline, yielding a score of $-9.27$~kcal/mol.

\section{Results}
\label{sec:results}

\subsection{Generation Statistics}

Vina-Direct guidance produced 9,997 valid molecules from 10,000 diffusion samples (99.97\% validity). The multi-objective strategy yielded 3,524 valid structures from 5,000 samples (70.5\%), with the lower rate attributable to more aggressive pocket-conditioning parameters. HNN-Denovo guidance produced 1,500 molecules across three independent runs. An unguided baseline of 4,000 molecules serves as the reference distribution. Property distributions across all four cohorts are shown in Figure~\ref{fig:boxplot}.

\subsection{Ablation Study: Guidance Strategy Comparison}

\begin{figure}[t]
\centering
\includegraphics[width=\linewidth]{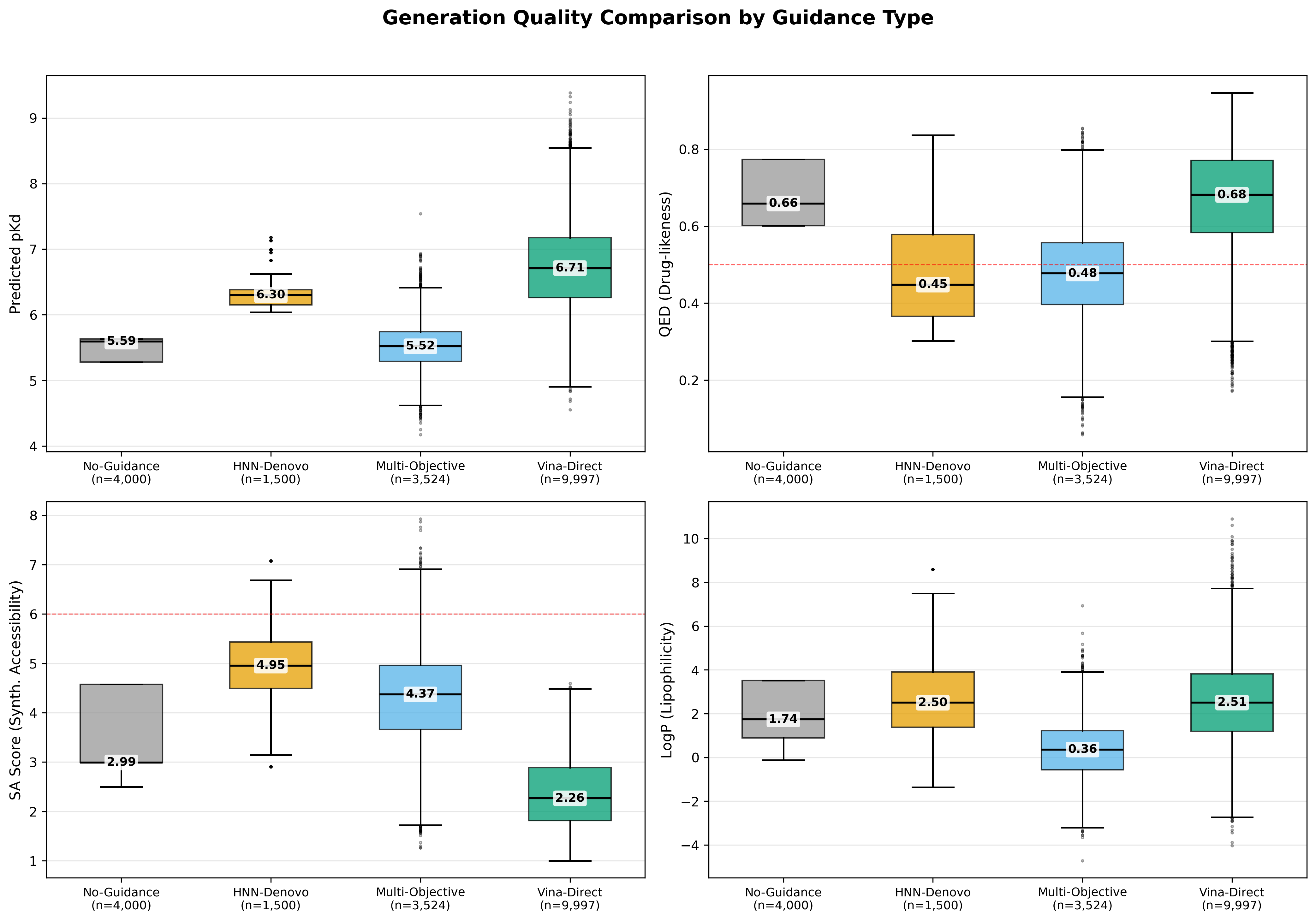}
\caption{Box-and-whisker comparison of molecular generation quality across four guidance strategies. Vina-Direct (green) achieves the highest binding affinity, best drug-likeness, and lowest synthetic complexity.}
\label{fig:boxplot}
\end{figure}

All 19,021 molecules across the four strategies were assessed on predicted binding affinity (pKd), drug-likeness (QED), synthetic accessibility (SA), and lipophilicity (LogP); box-and-whisker distributions are presented in Figure~\ref{fig:boxplot}.

Vina-Direct guidance dominates across all four metrics. Its median pKd of 6.71 exceeds HNN-Denovo (6.29) by 0.42 log units and Multi-Objective (5.52) by 1.19 log units. Critically, the Vina-Direct molecules are also the \textit{easiest to synthesize}: median SA of 2.26 versus 4.97 for HNN-Denovo and 4.37 for Multi-Objective. This counter-intuitive result—that stronger binders are more synthetically accessible—reflects the fact that Vina's physics-based scoring rewards compact, well-organized molecular geometries that happen to correlate with synthetic tractability, whereas learned proxies may favor structurally elaborate scaffolds to maximize predicted affinity.

The HNN-Denovo strategy shows a meaningful improvement over the unguided baseline in pKd ($+0.75$), confirming that even an approximate guidance signal biases generation toward active molecules. The HNN-Denovo model itself is accurate (PCC~$= 0.72$ on BindingDB), but the differentiable proxy score used to bridge its SMILES-based predictions into 3D coordinate gradients achieves only $r = 0.224$ correlation with Vina (Appendix~\ref{sec:proxy}). This domain mismatch---not a deficiency in the affinity predictor---limits the precision of the guidance bias, and the resulting SA scores (median 4.97) indicate that a distance-based proxy does not effectively select for synthetic tractability.

\subsection{Top Lead Compounds}

Table~\ref{tab:top_leads} presents the five top-ranked molecules after uniform redocking. All were generated under Vina-Direct guidance. The strongest binder achieves $-11.05$~kcal/mol (pKd~= 8.10), a 19.2\% improvement over the crystallographic reference of $-9.27$~kcal/mol (pKd~= 6.80). All five candidates exceed the reference by more than 1.2~kcal/mol, with zero Lipinski violations.

\begin{table*}[t]
\centering
\scriptsize
\begin{tabular}{clcccccc}
\toprule
\textbf{Rank} & \textbf{SMILES} & \textbf{Vina} & \textbf{pKd} & \textbf{SA} & \textbf{QED} & \textbf{MW} & \textbf{Lip.} \\
& & \textbf{(kcal/mol)} & & & & \textbf{(Da)} & \textbf{viol.}\\
\midrule
1 & \texttt{O=C(NC1CCNCC1c1ccc[nH]1)c1c(C2CCCOC2)ccc2ccccc12} & $-11.05$ & 8.10 & 3.79 & 0.618 & 403.5 & 0 \\
2 & \texttt{O=S(=O)(Cc1cccnc1)Nc1ccc(F)cc1-c1ccc2ccccc2c1C1CCNCC1} & $-10.62$ & 7.79 & 2.61 & 0.389 & 475.6 & 1 \\
3 & \texttt{NC(=O)NC(=O)CC1CNCCC1c1cccc2ccccc12} & $-10.59$ & 7.77 & 3.12 & 0.812 & 311.4 & 0 \\
4 & \texttt{NC(=O)Cc1c(NC(=O)c2cc(F)ccc2-c2cccnc2)ccc2ccccc12} & $-10.55$ & 7.74 & 2.26 & 0.524 & 399.4 & 0 \\
5 & \texttt{NC(=O)C1CNCCC1c1cc(F)ccc1-c1cnccc1C1CCCNC1} & $-10.51$ & 7.71 & 3.75 & 0.759 & 382.5 & 0 \\
\midrule
\textit{Ref.} & \textit{(crystallographic inhibitor)} & $-9.27$ & 6.80 & 3.34 & 0.680 & 370.5 & 0 \\
\bottomrule
\end{tabular}
\caption{Top five generated molecules ranked by Vina docking score after uniform redocking against ACVR1 (PDB: 3MTF). All surpass the crystallographic reference by $>$1.2~kcal/mol. Molecule~3 (MW~311, QED~0.81, SA~3.1) represents a particularly strong lead: high affinity, high drug-likeness, and low synthetic complexity.}
\label{tab:top_leads}
\end{table*}

Molecule~3 (rank~3, $-10.59$~kcal/mol) is notable for its combination of high affinity, high QED (0.812), and low molecular weight (311~Da), making it the most balanced lead candidate. Its urea-linked naphthyl-piperidine scaffold suggests pharmacophore features compatible with kinase hinge-binding motifs.

\begin{figure}[t]
\centering
\includegraphics[width=\linewidth]{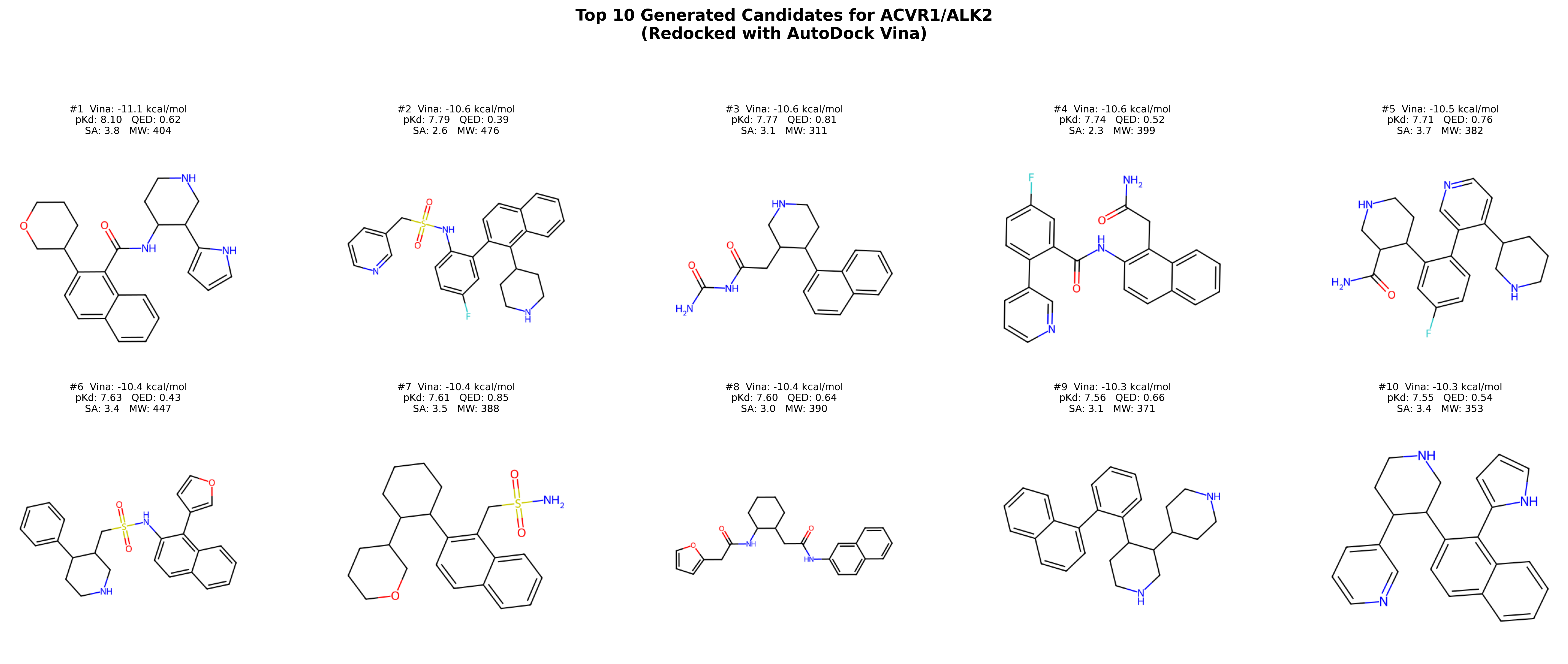}
\caption{2D structures of the top 10 generated ACVR1 inhibitor candidates with docking scores and drug-likeness properties.}
\label{fig:structures}
\end{figure}

2D structures of the top 10 candidates are shown in Figure~\ref{fig:structures}.

\begin{figure}[t]
\centering
\includegraphics[width=0.5\linewidth]{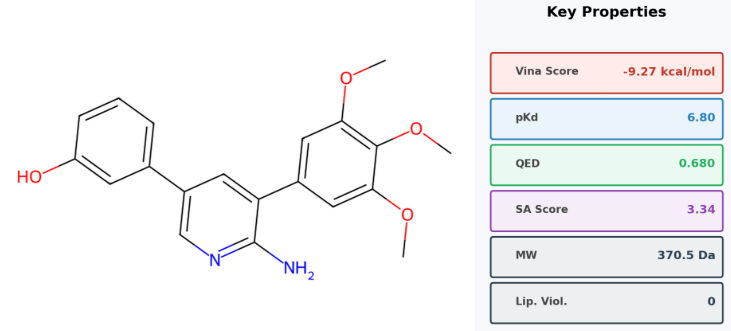}
\caption{2D chemical structure of the 3MTF crystallographic reference inhibitor (A3F). This baseline ligand established the benchmark for evaluating all generated candidates.}
\label{fig:reference}
\end{figure}

\subsection{Drug-Likeness and Property Distributions}

\begin{figure}[t]
\centering
\includegraphics[width=\linewidth]{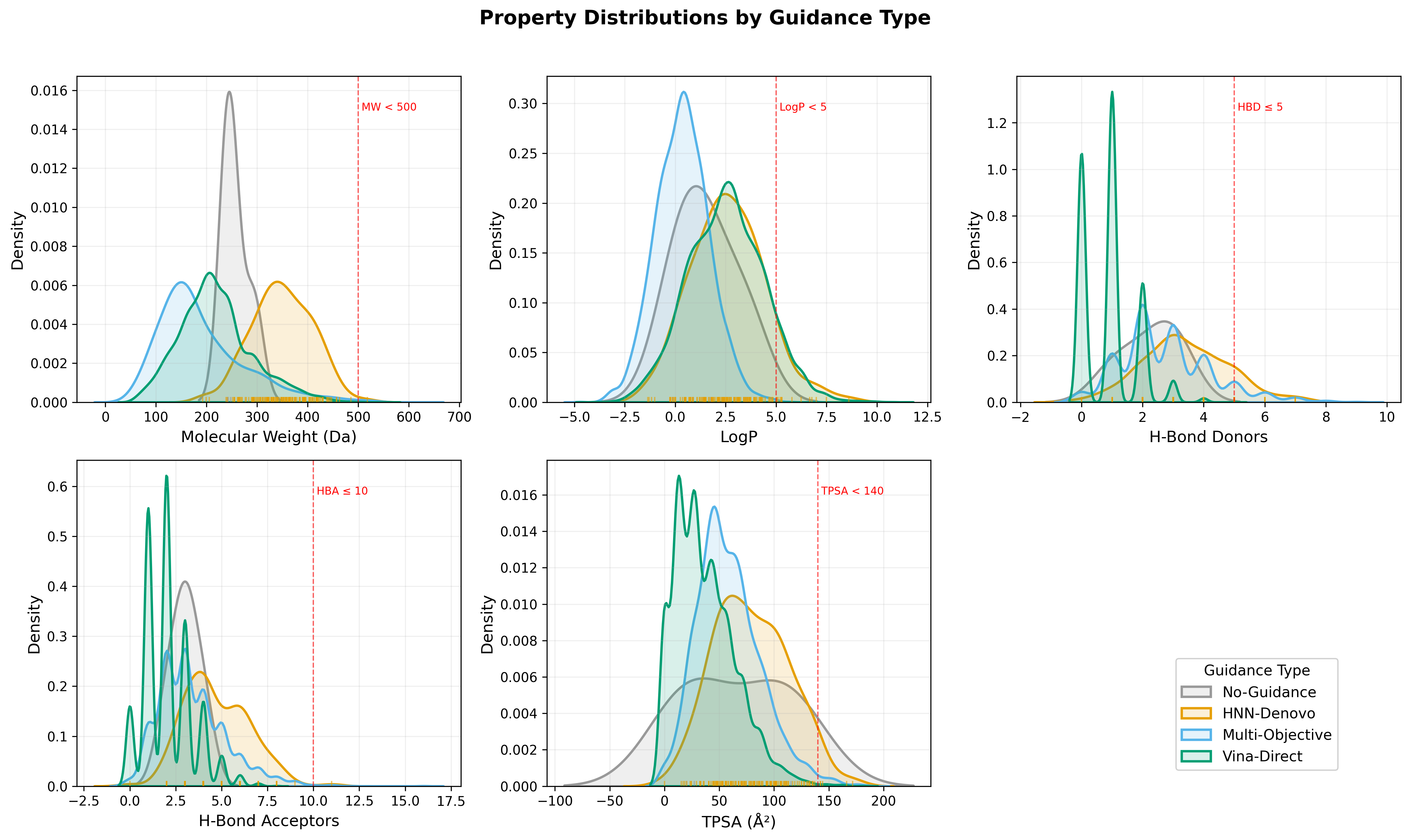}
\caption{Property distributions by guidance type. Lipinski reference thresholds are shown as red dashed lines. Vina-Direct molecules (green) cluster within drug-like space across all five properties.}
\label{fig:distributions}
\end{figure}

Among the top 100 redocked molecules (Vina scores $-11.05$ to $-9.13$~kcal/mol), all 100 are Lipinski-compliant (zero violations). The median SA score of 2.67 indicates that most top candidates are synthetically tractable—a marked improvement over the multi-objective route (median SA~= 4.37). Property distributions across all 19,021 molecules are shown in Figure~\ref{fig:distributions}.

\begin{figure}[t]
\centering
\includegraphics[width=\linewidth]{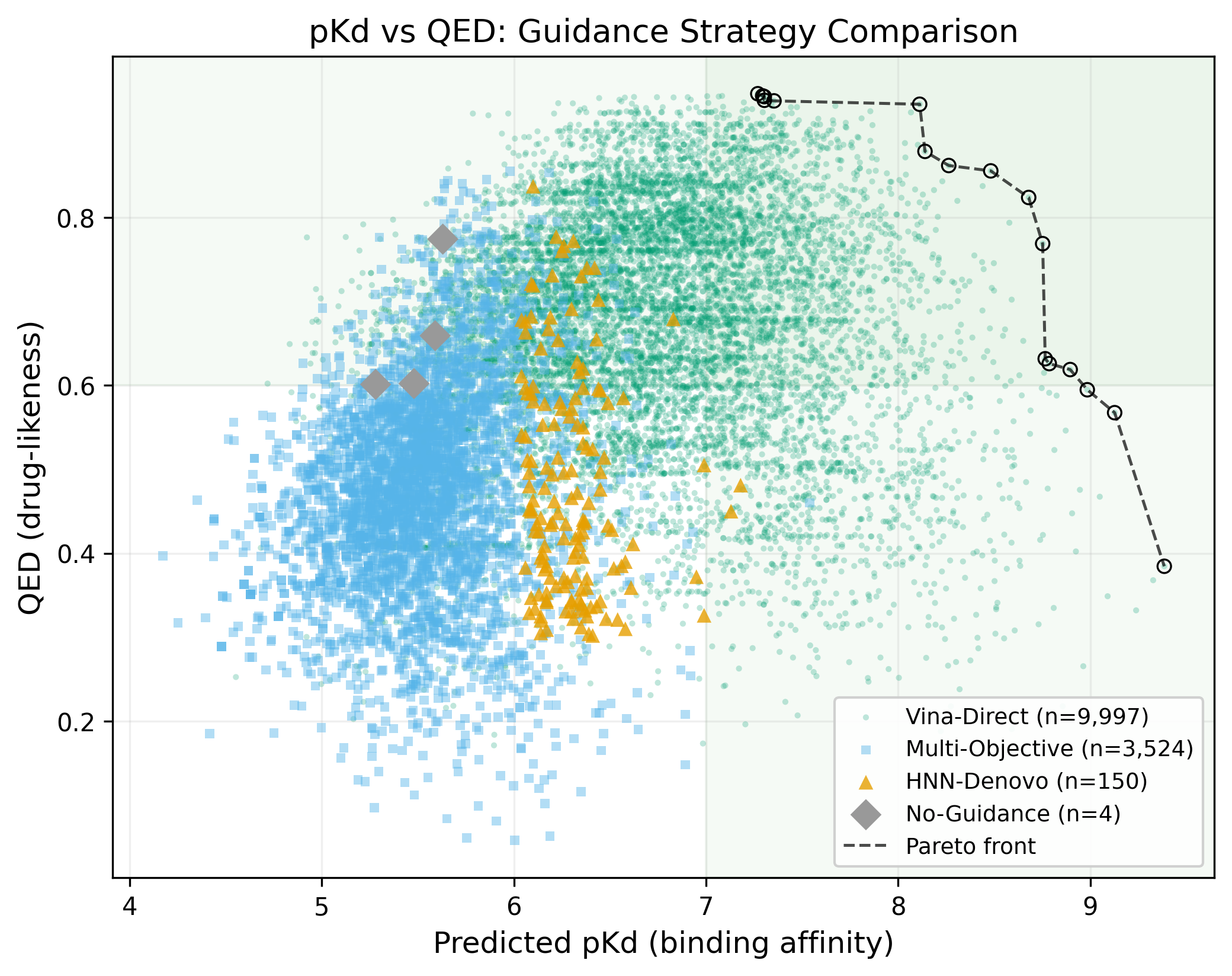}
\caption{pKd vs QED across all 19,021 molecules. The Pareto front (dashed line) traces the non-dominated trade-off between binding affinity and drug-likeness. Vina-Direct molecules (green) dominate the frontier.}
\label{fig:pareto}
\end{figure}

The pKd versus QED trade-off is visualized in Figure~\ref{fig:pareto}, which shows the Pareto frontier across all 19,021 molecules. The Vina-Direct strategy occupies the favorable upper-right quadrant (high affinity, high drug-likeness), with the Pareto front composed predominantly of Vina-guided molecules.

\subsection{Chemical Diversity}

\begin{figure}[t]
\centering
\includegraphics[width=\linewidth]{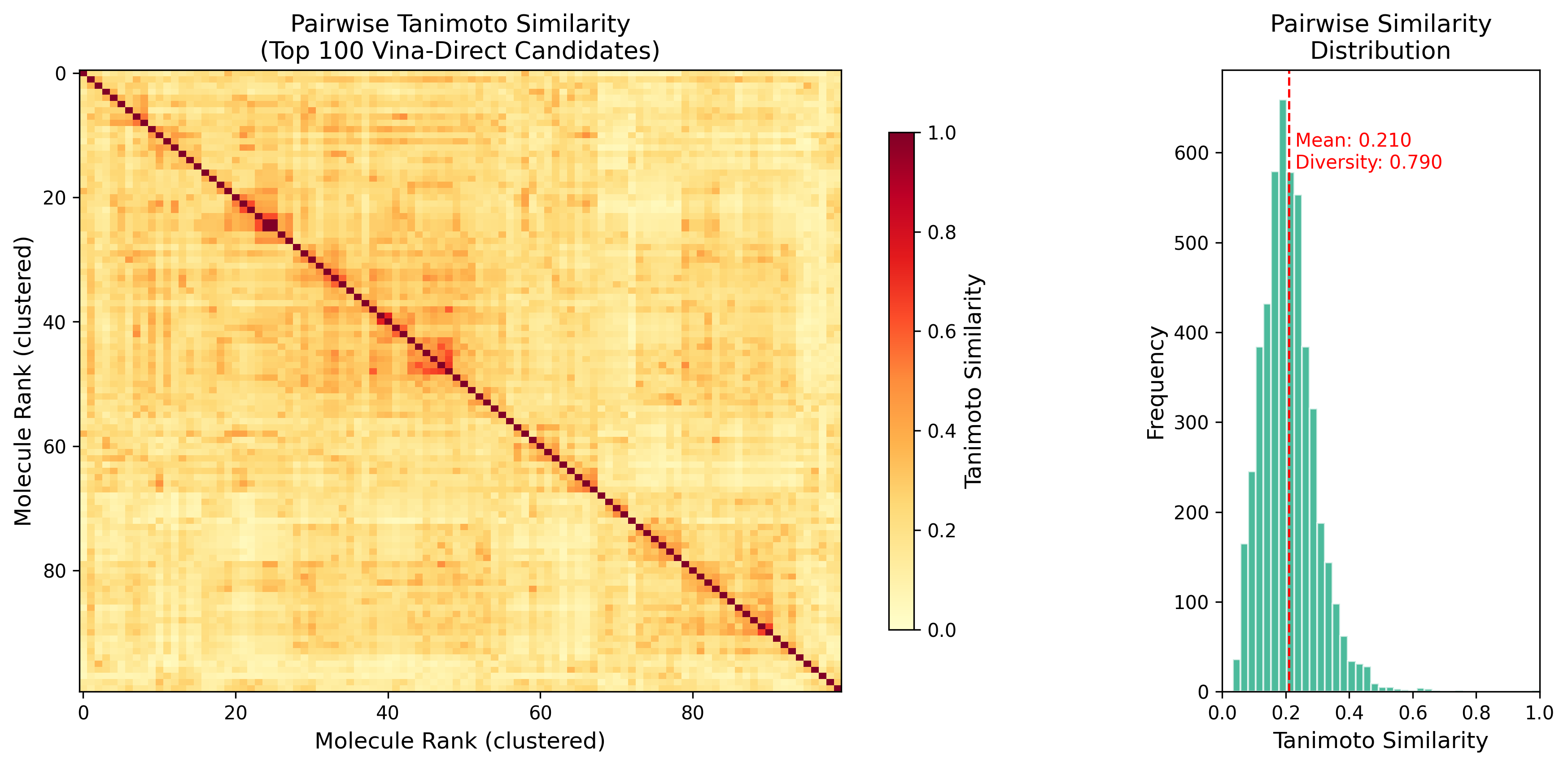}
\caption{Chemical diversity of the top 100 candidates. Left: pairwise Tanimoto similarity heatmap (hierarchically clustered). Right: distribution of off-diagonal similarities. Mean Tanimoto = 0.210 indicates high structural diversity.}
\label{fig:diversity}
\end{figure}

A key concern with guided generation is mode collapse—whether the guidance signal forces all molecules toward a single scaffold. To assess this, we computed pairwise Tanimoto similarity (Morgan fingerprints, radius~2, 2048 bits) among the top 100 redocked candidates (Figure~\ref{fig:diversity}).

The mean pairwise Tanimoto similarity is 0.210, yielding an internal diversity of $1 - 0.210 = 0.790$. For reference, random drug-like molecules typically exhibit pairwise similarities of $0.15$--$0.25$, placing our top candidates squarely within the expected range for a structurally diverse set. Hierarchical clustering reveals several distinct scaffold families among the top 100, with no single cluster dominating, confirming that Vina gradient guidance produces diverse solutions rather than collapsing onto a single binding mode.

\section{Discussion}

\subsection{Summary of Contributions}

This work demonstrates three key findings: (1) real-time Vina gradient injection into GCDM produces ACVR1 inhibitors 19.2\% stronger than the crystallographic reference while maintaining Lipinski compliance and low synthetic complexity; (2) physics-based docking guidance (Vina-Direct) dominates learned proxies (HNN-Denovo) and multi-objective scoring across all metrics; and (3) the process maintains chemical diversity—top candidates span multiple scaffold families (internal diversity 0.790).

\subsection{Vina-Direct versus HNN-Denovo}

The HNN-Denovo guidance signal offers a meaningful speedup ($\sim$60$\times$ per guidance step) and improves median pKd over the 4,000-molecule unguided baseline by 0.75 log units across 1,500 generated candidates. The underlying affinity predictor is accurate (PCC~$= 0.72$, RMSE~$= 0.70$, MAE~$= 0.53$ pKd on BindingDB), confirming it can distinguish potent from weak binders in sequence--ligand space. The performance gap relative to Vina-Direct stems not from model inaccuracy but from the proxy scoring function that must approximate HNN-Denovo's predictions in 3D coordinate space during diffusion; this proxy achieves only $r = 0.224$ against Vina ground truth. For rapid early-stage exploration of a new target---where coarse directional guidance suffices---HNN-Denovo with its proxy remains a practical choice. For final candidate generation against well-characterized targets like ACVR1, where gradient accuracy is critical, Vina-Direct guidance is essential.

\subsection{Limitations}

Several limitations merit explicit acknowledgment. First, all binding evaluations rely on rigid-receptor docking with AutoDock Vina, which does not capture induced-fit effects, solvation dynamics, or entropic contributions to binding. Second, no ADMET (absorption, distribution, metabolism, excretion, toxicity) or off-target kinase selectivity assessment was performed; these factors are often decisive failure points in kinase inhibitor development. Third, Vina gradient guidance is computationally expensive: each guidance step requires $6N+1$ docking evaluations per molecule, amounting to several seconds per molecule per step. Fourth, dissociation kinetics are estimated from an empirical pKd-$k_\text{off}$ correlation rather than from molecular dynamics simulations. Finally, this work constitutes a computational proof of concept; experimental synthesis and biophysical validation of top candidates remain necessary steps toward therapeutic translation.

\subsection{Future Directions}

Future extensions should prioritize experimental validation of top candidates through biochemical ALK2 activity assays and cellular ossification models. Computational improvements include replacing the empirical kinetics model with molecular-dynamics-based residence-time estimation, incorporating ADMET predictors directly into the guidance signal, and applying active learning to iteratively refine the generative model based on experimental feedback. The modular design of KinetiDiff—separating the diffusion backbone from the guidance oracle—enables straightforward substitution of Vina with more accurate but computationally expensive scoring functions (e.g., MM-GBSA or free-energy perturbation) as GPU-accelerated implementations become available.

\section{Conclusion}

We have presented KinetiDiff, a framework for generating kinase inhibitor candidates by injecting AutoDock Vina docking gradients into the reverse diffusion process of a Geometry-Complete Diffusion Model. Applied to ACVR1—the causative kinase in FOP—the framework produces molecules with binding affinities up to $-11.05$~kcal/mol, a 19.2\% improvement over the crystallographic reference, while maintaining full drug-likeness compliance and high structural diversity. Systematic ablation demonstrates the superiority of real-time docking guidance over learned proxy signals and unguided generation. These results establish gradient-guided geometric diffusion as a practical computational tool for rare-disease drug discovery, providing a foundation for subsequent experimental validation and clinical translation.

\section*{Acknowledgments}
We gratefully acknowledge the computational resources provided by Google's TPU Research Cloud \citep{GoogleTRC2023} and Vast.ai \citep{VastAI2023}.
We thank Dr.~Ian~S.~Haworth of the USC Mann School of Medicine/Pharmacology for his valuable guidance on computational approaches to this project.
We also thank Frederick Kaplan of the University of Pennsylvania Perelman School of Medicine, Robert Pignolo of the Mayo Clinic, and Dr.~Patricia Delai of the International Fibrodysplasia Ossificans Progressiva Association (IFOPA) for their insightful discussions and domain expertise on heterotopic ossification that helped shape this research.

{\setstretch{1.5}
\bibliographystyle{unsrtnat}
\bibliography{references}
}

\clearpage
\appendix

\section{Computational Methods and Hyperparameters}

\subsection{GCDM Architecture}

The equivariant denoising network comprised 6 GCP message-passing layers with hidden dimension 256 and SiLU activations. Graphs were constructed from ligand atoms and pocket residues within a 10~\AA{} cutoff radius, with edges formed between all atom pairs within this radius. Edge features consisted of 50 Gaussian RBF kernels (centers evenly spaced 0--10~\AA, width $\gamma = 0.5$~\AA$^{-1}$). Table~\ref{tab:gcdm-hyperparams} provides the complete GCPNet architecture specification.

\begin{table}[h]
\centering
\small
\caption{GCPNet denoising network hyperparameters.}
\label{tab:gcdm-hyperparams}
\begin{tabular}{ll}
\toprule
\textbf{Parameter} & \textbf{Value} \\
\midrule
GCP interaction layers & 6 \\
Node scalar hidden dim ($d_s$) & 256 \\
Node vector hidden dim ($d_v$) & 32 \\
Edge scalar hidden dim & 16 \\
Edge vector hidden dim & 8 \\
Scalar nonlinearity & SiLU \\
Vector nonlinearity & SiLU \\
Bottleneck factor & 4 \\
Vector gating & Enabled \\
Scalar message attention & Enabled \\
Joint feature space dim & 32 \\
Dropout & 0.0 \\
Ligand atom types & 9 (C, N, O, F, P, S, Cl, Br, I) \\
Pocket residue types & 20 (standard amino acids) \\
RBF kernels (edge features) & 50 (0--10~\AA, $\gamma = 0.5$~\AA$^{-1}$) \\
\bottomrule
\end{tabular}
\end{table}

\subsection{Diffusion Schedule}

The forward diffusion process used $T = 1000$ timesteps. Training employed a polynomial noise schedule:
\begin{equation}
\bar{\alpha}_t = \max\!\left(\left(1 - \left(\frac{t}{T}\right)^2\right)^2,\; 10^{-3}\right),
\label{eq:polynomial-schedule}
\end{equation}
while inference used the cosine schedule (Eq.~\ref{eq:cosine-schedule}). Sampling was accelerated via DDIM with 50 steps (stride 20, $\eta = 0$), with coordinate clipping to $[-10, 10]$~\AA{} at each step to prevent numerical drift. Noise precision was set to $10^{-5}$. Coordinate and atom-feature normalization factors were 1 and 4, respectively.

\subsection{Docking and Guidance Parameters}

All redocking experiments used AutoDock Vina~1.2.3 against ACVR1 wild-type (PDB: 3MTF). The receptor was protonated at pH~7.4 and assigned Gasteiger charges. The search box was $22 \times 22 \times 22$~\AA{} centered at the reference-ligand centroid $(24.87, -12.54, 38.40)$~\AA. All runs used seed~42. 3D conformers were embedded using RDKit's ETKDGv3 (with v2 and random-coordinate fallbacks), minimized with MMFF94 (UFF fallback), and converted to PDBQT format via Meeko's \texttt{MoleculePreparation} and \texttt{PDBQTWriterLegacy}. Table~\ref{tab:vina-hyperparams} summarizes the full Vina guidance configuration.

\begin{table}[h]
\centering
\small
\caption{Vina gradient guidance hyperparameters.}
\label{tab:vina-hyperparams}
\begin{tabular}{ll}
\toprule
\textbf{Parameter} & \textbf{Value} \\
\midrule
Finite-difference $\varepsilon$ & 0.05~\AA \\
Base guidance scale $\lambda_0$ & 0.1 \\
$\lambda$ clipping range & $[0.1,\, 10.0]$ \\
Guidance active window & $0.2T$ to $T$ (last 80\%) \\
Guidance application interval & Every 10 denoising steps \\
Gradient norm clipping & 10.0 (then rescale to 1.0) \\
Exhaustiveness (guidance) & 1 \\
Exhaustiveness (validation) & 4 \\
Vina docking mode (guidance) & \texttt{--local\_only} \\
Docking box & $22 \times 22 \times 22$~\AA \\
Docking box center & $(24.87, -12.54, 38.40)$~\AA \\
Vina timeout per call & 60~s \\
\bottomrule
\end{tabular}
\end{table}

\subsection{Training Data and HNN-Denovo Architecture}

The HNN-Denovo affinity predictor was trained on BindingDB \citep{Liu2007BindingDB} (November 2024 release, $\sim$2.8M entries), filtered to ACVR1-family kinases (ACVR1, ACVR1B, ACVR1C, BMPR1A/B, BMPR2). Only numeric affinity measurements ($K_d$, $K_i$, IC$_{50}$, EC$_{50}$) within 0.1~nM--100~$\mu$M were retained. The final dataset contained 15,324 protein--ligand pairs (2,847 ACVR1-specific). Data were split 70/15/15 for training/validation/test. Models were trained for 50 epochs with Adam ($\text{lr} = 10^{-4}$, weight decay $10^{-5}$), batch size 32, and early stopping (patience 10).

\paragraph{HNN-Denovo Architecture.}
The affinity predictor used parallel 1D convolutional encoders (kernels $k \in \{3, 5, 7\}$, 256 channels each) for ligand SMILES (max length 200) and protein sequences (max length 1000, 22-token vocabulary: 20 amino acids + unknown + padding). Embedding dimension was 64, hidden dimension 256, with 3 convolutional layers and dropout 0.1. Affinity predictions were normalized using BindingDB training-set statistics ($\mu = 6.46$, $\sigma = 1.38$ pKd).

\paragraph{HNN-Denovo Performance.}
On the held-out BindingDB test set (70,248 training samples), HNN-Denovo achieves PCC~$= 0.72$, RMSE~$= 0.70$~pKd, and MAE~$= 0.53$~pKd---comparable to published SMILES-based affinity predictors. Note that this accuracy characterizes the model's prediction of pKd from sequence--ligand pairs; the lower correlation observed against Vina in the guidance context ($r = 0.224$, Appendix~\ref{sec:proxy}) reflects the limitations of the differentiable 3D proxy used to bridge HNN-Denovo's SMILES-based domain into the coordinate space of the diffusion loop, and should not be interpreted as a measure of HNN-Denovo's intrinsic quality.

\subsection{Multi-Objective Scoring Parameters}

The adaptive multi-objective scoring system (Section~\ref{sec:scoring}) used the objectives and thresholds summarized in Table~\ref{tab:multi-obj-params}.

\begin{table}[h]
\centering
\small
\caption{Multi-objective scoring targets and thresholds.}
\label{tab:multi-obj-params}
\begin{tabular}{llll}
\toprule
\textbf{Objective} & \textbf{Target} & \textbf{Threshold} & \textbf{Direction} \\
\midrule
Affinity (pKd) & 7.0 & $\geq 6.0$ & Maximize \\
Docking (kcal/mol) & $-11.0$ & $\leq -10.0$ & Minimize \\
SA score & 3.0 & $\leq 3.5$ & Minimize \\
\bottomrule
\end{tabular}
\end{table}

\noindent Adaptive threshold weights were updated with rate $\alpha = 0.1$ and bounded to $[0.1, 0.6]$. The SA objective enforced a minimum weight floor of 0.25 and an exponential penalty multiplier of 10.0 for SA scores exceeding the threshold. Final composite ranking used weights: affinity 0.35, docking 0.30, SA 0.25, QED 0.10.

\subsection{Computational Resources}

Generation experiments were conducted on dual NVIDIA RTX A6000 GPUs (49~GB VRAM each) with an AMD EPYC 7543 CPU and 512~GB RAM. Vina-Direct guidance required approximately 2.3~hours per 2,000 molecules per GPU. Uniform redocking of 3,000 pre-ranked candidates across 10 parallel workers completed in 36~minutes.

\section{Docking Visualization}
\label{appendix:docking-viz}

\begin{figure}[h]
\centering
\begin{subfigure}[t]{0.46\columnwidth}
  \centering
  \includegraphics[width=\linewidth]{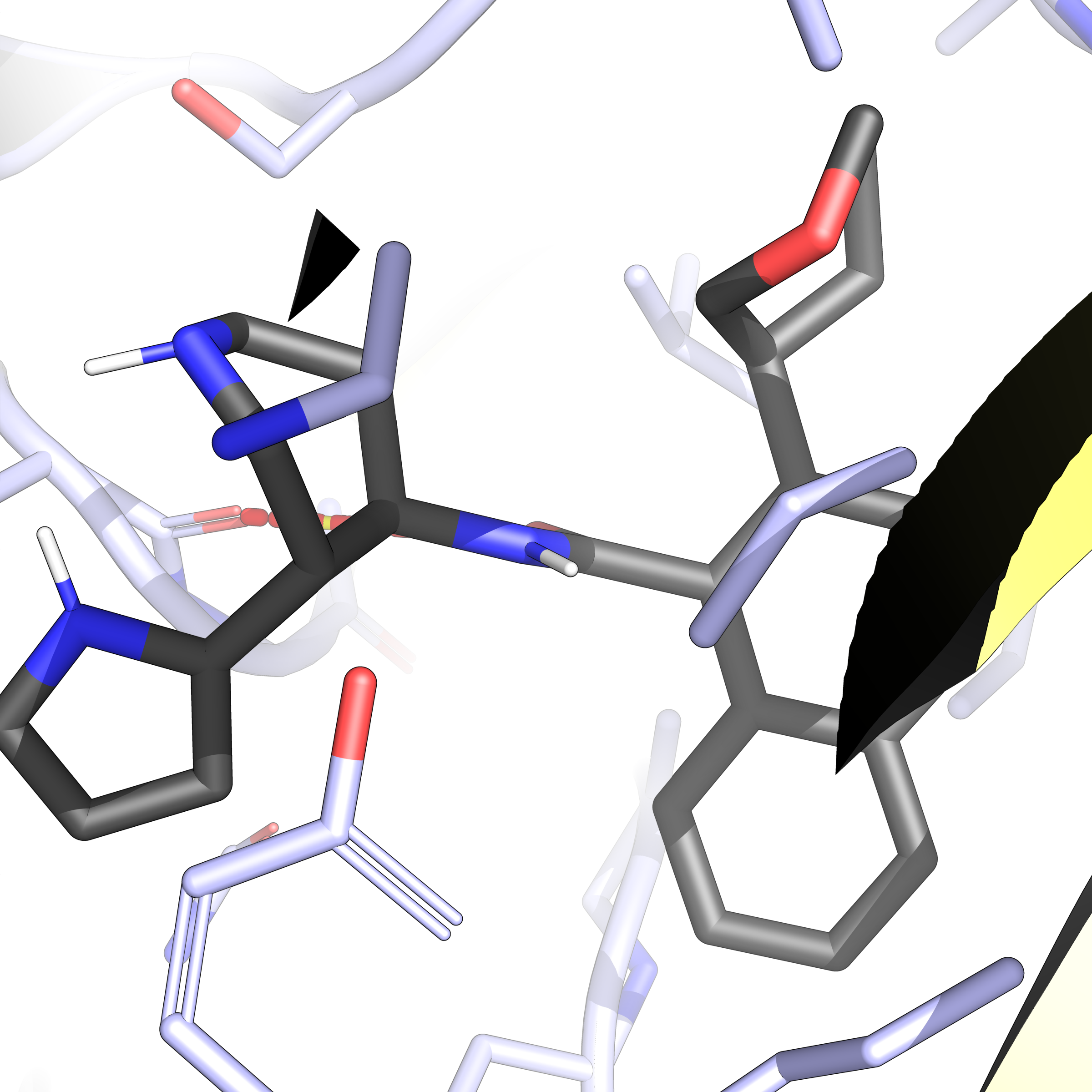}
  \caption{Rank 1 (Vina: $-11.05$~kcal/mol). 10 pocket residues, 1 H-bond. MW~403.5~Da, QED~0.618.}
  \label{fig:docking-rank1}
\end{subfigure}
\hfill
\begin{subfigure}[t]{0.46\columnwidth}
  \centering
  \includegraphics[width=\linewidth]{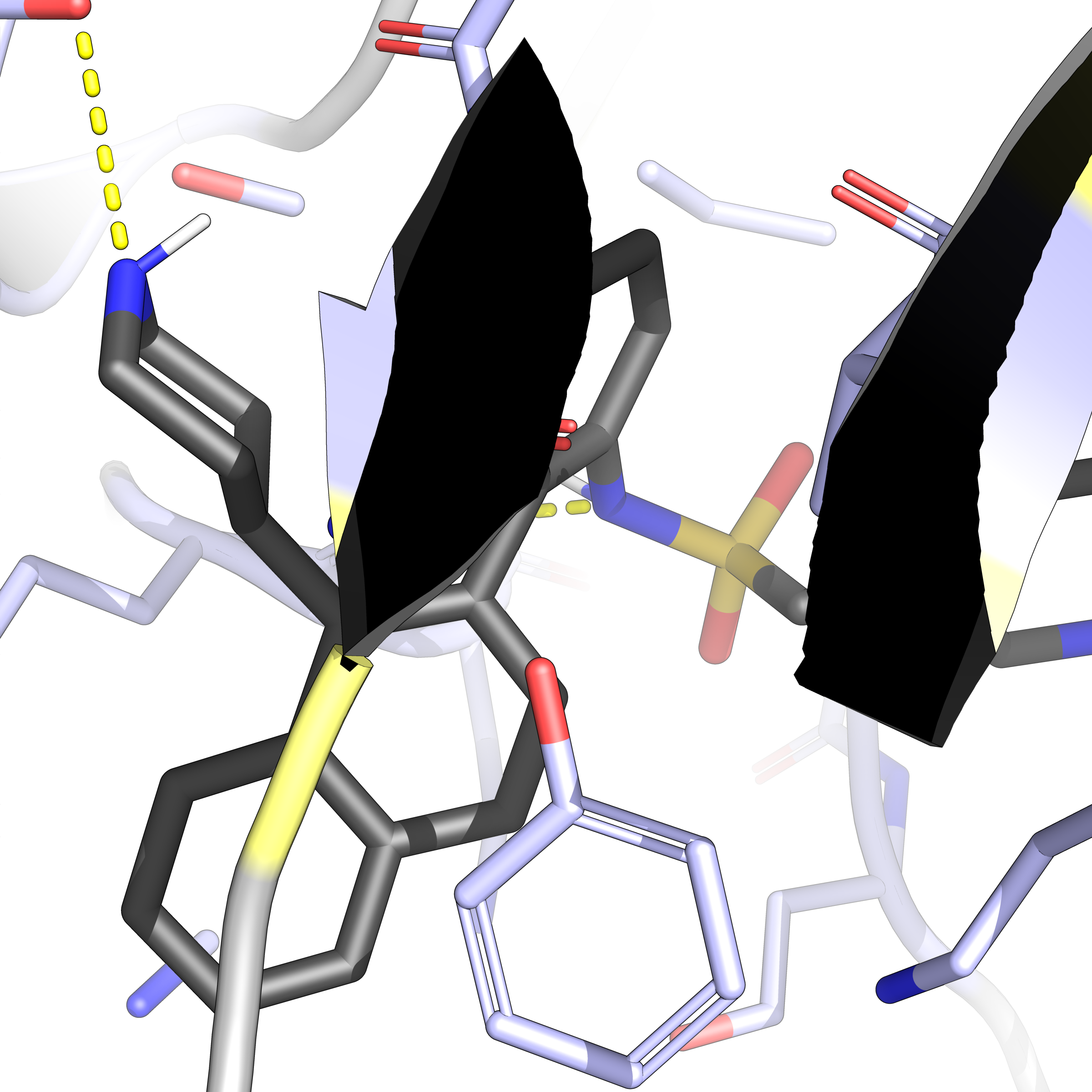}
  \caption{Rank 2 (Vina: $-10.62$~kcal/mol). 8 pocket residues, 2 H-bonds. MW~475.6~Da, QED~0.389.}
  \label{fig:docking-rank2}
\end{subfigure}
\caption{Top two generated molecules docked in the ACVR1 binding pocket (PDB: 3MTF), rendered in PyMOL. Protein backbone shown in cartoon (light blue); pocket residues and ligand as sticks; yellow dashes indicate hydrogen bonds.}
\label{fig:docking-poses}
\end{figure}

Figure~\ref{fig:docking-poses} shows the top two generated molecules docked in the ACVR1 binding pocket (PDB: 3MTF), rendered in PyMOL \citep{PyMOL2024}. Docking was performed with AutoDock Vina \citep{Trott2010AutoDock}. Protein backbone is shown in cartoon representation (light blue), binding pocket residues within 5~\AA{} of the ligand are shown as sticks, and the ligand is rendered as colored sticks. Yellow dashed lines indicate hydrogen bonds.

\clearpage
\section{Sampling Algorithm}
\label{appendix:algorithm}

\begin{algorithm}[h]
\caption{KinetiDiff: Vina-Guided Reverse Diffusion Sampling}
\label{alg:kinetidiff}
\begin{algorithmic}[1]
\Require Protein pocket $\mathcal{P}$, diffusion model $\boldsymbol{\epsilon}_\theta$, Vina scorer $\mathcal{U}_{\text{Vina}}$, steps $T$, schedule $\lambda_t$ (Eq.~\ref{eq:guidance-schedule})
\Ensure Generated ligand $\mathbf{x}_0$

\State Sample $\mathbf{z}_T = (\mathbf{x}_T, \mathbf{h}_T) \sim \mathcal{N}(\mathbf{0}, \mathbf{I})$ \Comment{$\mathbf{x}_T \in \mathbb{R}^{N\times 3},\; \mathbf{h}_T \in \mathbb{R}^{N\times 9}$}

\For{$t = T, T-1, \dots, 1$}
    \State $\hat{\mathbf{x}}_0^{(t)} \gets \text{Denoise}(\mathbf{x}_t, \mathcal{P}, t;\, \boldsymbol{\epsilon}_\theta)$ \Comment{GCDM denoising step}

    \If{$t \geq 0.2T$ \textbf{and} $t \bmod 10 = 0$} \Comment{Guidance activation condition}
        \State Convert $\mathbf{x}_t$ to 3D conformer and PDBQT format
        \For{each atom $i = 1, \dots, N$;\; dimension $d \in \{x,y,z\}$}
            \State $\hat{g}_i^d \gets \frac{\mathcal{U}_{\text{Vina}}(\mathbf{x}_t+\varepsilon\,\mathbf{e}_{i,d}) - \mathcal{U}_{\text{Vina}}(\mathbf{x}_t-\varepsilon\,\mathbf{e}_{i,d})}{2\varepsilon}$ \Comment{$\varepsilon = 0.05$~\AA}
        \EndFor
        \If{$\|\hat{\mathbf{g}}_t\| > 10$} \Comment{Gradient norm clipping}
            \State $\hat{\mathbf{g}}_t \gets \hat{\mathbf{g}}_t \cdot 10\,/\,\|\hat{\mathbf{g}}_t\|$
        \EndIf
        \State $\hat{\mathbf{g}}_t \gets \hat{\mathbf{g}}_t \,/\, (\|\hat{\mathbf{g}}_t\| + 10^{-8})$ \Comment{Normalize to unit length}
        \State $\mathbf{x}_{t-1} \gets \hat{\mathbf{x}}_0^{(t)} - \lambda_t \cdot \hat{\mathbf{g}}_t$ \Comment{Guided update (Eq.~\ref{eq:guided-update})}
    \Else
        \State $\mathbf{x}_{t-1} \gets \hat{\mathbf{x}}_0^{(t)}$ \Comment{Unguided denoising step}
    \EndIf
    \State $\mathbf{x}_{t-1} \gets \mathbf{x}_{t-1} - \frac{1}{N}\sum_{i}\mathbf{x}_{t-1,i}$ \Comment{Re-center to zero CoM}
\EndFor
\State Discretize atom types $\mathbf{h}_0$ from continuous predictions \Comment{Eq.~\ref{eq:atom-discretization}}
\State \Return $(\mathbf{x}_0, \mathbf{h}_0)$
\end{algorithmic}
\end{algorithm}

\clearpage
\section{ACVR1 Biology and Inhibition Mechanism}
\label{appendix:biology}

\begin{figure}[H]
\centering
\includegraphics[width=\textwidth]{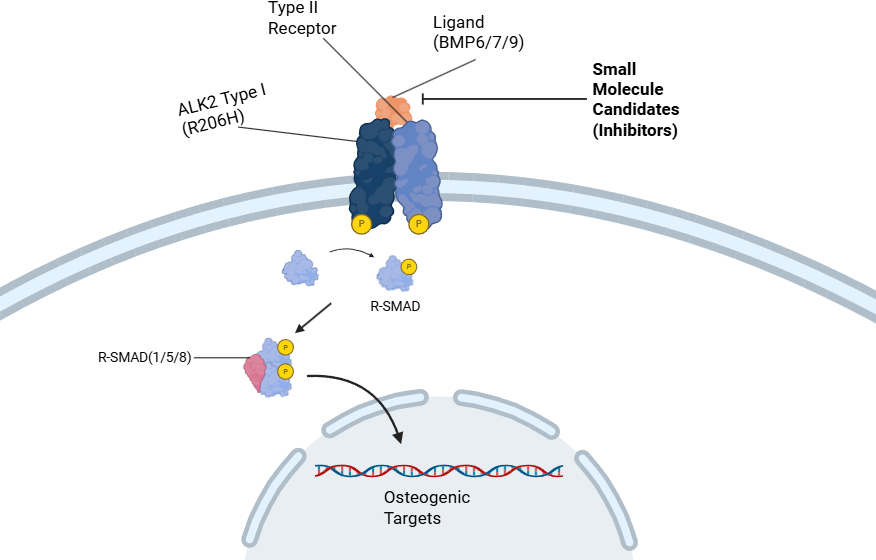}
\caption{ACVR1/ALK2 signaling pathway and mechanism of heterotopic ossification in FOP. \textbf{Left:} Normal BMP signaling through wild-type ALK2, which is activated only by BMP ligands and remains unresponsive to activin~A. \textbf{Center:} The R206H gain-of-function mutation renders ALK2 constitutively active and responsive to activin~A, driving aberrant SMAD1/5/8 phosphorylation and downstream osteogenic differentiation. \textbf{Right:} Small-molecule inhibition at the ATP-binding site; transient inhibition (fast $k_\text{off}$) suppresses pathological flares while permitting recovery of normal BMP signaling between episodes. Created with BioRender \citep{BioRender2026}.}
\label{fig:biology}
\end{figure}

%

\clearpage
\section{Detailed Mathematical Formulations}
\label{appendix:formulas}

This appendix provides complete mathematical specifications for all components of the KinetiDiff framework, including expanded derivations, noise schedules, and implementation-level details.

\subsection{Forward Diffusion Process (Expanded)}

The forward marginal distribution can equivalently be written in the variance-preserving (VP) parameterization as:
\begin{equation}
q(\mathbf{x}_t \mid \mathbf{x}_0) = \mathcal{N}\!\left(\mathbf{x}_t;\, \alpha_t \mathbf{x}_0,\, \sigma_t^2 \mathbf{I}\right),
\label{eq:forward-vp}
\end{equation}
where $\alpha_t = \sqrt{\bar{\alpha}_t}$ is the signal coefficient and $\sigma_t = \sqrt{1 - \bar{\alpha}_t}$ is the noise coefficient, with $\bar{\alpha}_t$ defined by the chosen noise schedule. The log signal-to-noise ratio is $\gamma_t = \log(\sigma_t^2 / \alpha_t^2)$.

\paragraph{Denoising Prediction.}
Given the noise prediction $\hat{\boldsymbol{\epsilon}}_t = \boldsymbol{\epsilon}_\theta(\mathbf{z}_t, t)$ from the GCPNet, the predicted clean sample is recovered via:
\begin{equation}
\hat{\mathbf{x}}_0 = \frac{1}{\alpha_t}\left(\mathbf{z}_t - \sigma_t\, \hat{\boldsymbol{\epsilon}}_t\right),
\label{eq:denoising-prediction}
\end{equation}
where:
\begin{itemize}[nosep,leftmargin=1.5em]
    \item $\mathbf{z}_t$ -- noised latent state at timestep $t$
    \item $\hat{\boldsymbol{\epsilon}}_t$ -- neural network's noise prediction
    \item $\alpha_t, \sigma_t$ -- signal and noise coefficients from the schedule
\end{itemize}

\paragraph{Reverse Step Mean.}
The posterior mean for the reverse transition from timestep $t$ to $s < t$ is:
\begin{equation}
\boldsymbol{\mu}_{t \to s} = \frac{1}{\alpha_{t|s}} \mathbf{z}_t - \frac{\sigma_{t|s}^2}{\sigma_t \cdot \alpha_{t|s}}\, \hat{\boldsymbol{\epsilon}}_t,
\label{eq:reverse-step-mean}
\end{equation}
with reverse-step sampling:
\begin{equation}
\mathbf{z}_s \sim \mathcal{N}\!\left(\boldsymbol{\mu}_{t \to s},\; \frac{\sigma_{t|s} \cdot \sigma_s}{\sigma_t}\, \mathbf{I}\right),
\end{equation}
where:
\begin{itemize}[nosep,leftmargin=1.5em]
    \item $\alpha_{t|s} = \alpha_t / \alpha_s$ -- transition ratio between steps
    \item $\sigma_{t|s}^2 = -\mathrm{expm1}(\mathrm{softplus}(\gamma_s) - \mathrm{softplus}(\gamma_t))$ -- transition noise variance
\end{itemize}

\subsection{Noise Schedules}

\paragraph{Cosine Schedule (Inference).}
As defined in Eq.~\ref{eq:cosine-schedule} with offset $s = 0.008$, providing smooth signal decay from data to noise.

\paragraph{Polynomial Schedule (Training).}
\begin{equation}
\bar{\alpha}_t = \max\!\left(\left(1 - \left(\frac{t}{T}\right)^2\right)^{\!2},\; 10^{-3}\right),
\label{eq:poly-schedule-appendix}
\end{equation}
which provides a smoother noise injection profile during training compared to cosine. The clipping at $10^{-3}$ prevents numerical instability at $t \approx T$.

\subsection{Guidance Injection Details}

\paragraph{Full Guided Update.}
The complete guidance-modified reverse step is:
\begin{equation}
\mathbf{x}_{t-1} = \hat{\mathbf{x}}_0^{(t)} - \lambda_t\, \hat{\mathbf{g}}_t,
\end{equation}
where $\hat{\mathbf{x}}_0^{(t)}$ is the GCDM denoiser's predicted clean-data estimate at step $t$, $\lambda_t$ follows the schedule in Eq.~\ref{eq:guidance-schedule}, and $\hat{\mathbf{g}}_t$ is the normalized and clipped numerical gradient (Eq.~\ref{eq:finite-diff}).

\paragraph{Gradient Clipping Protocol.}
Applied in two stages to prevent destabilizing updates:
\begin{enumerate}[nosep]
    \item \textbf{Norm clipping:} If $\|\hat{\mathbf{g}}_t\| > 10.0$, rescale: $\hat{\mathbf{g}}_t \leftarrow \hat{\mathbf{g}}_t \cdot 10.0 \,/\, \|\hat{\mathbf{g}}_t\|$.
    \item \textbf{Post-scaling normalization:} After applying $\lambda_t$, if $\|\lambda_t \hat{\mathbf{g}}_t\| > 1.0$, normalize: $\lambda_t \hat{\mathbf{g}}_t \leftarrow \lambda_t \hat{\mathbf{g}}_t \,/\, \|\lambda_t \hat{\mathbf{g}}_t\|$.
\end{enumerate}
The finite-difference step size is $\varepsilon = 0.05$~\AA. Each gradient evaluation requires $6N + 1$ Vina scoring calls.

\subsection{Center-of-Mass Constraint}

After every reverse diffusion step, coordinates are re-centered to the zero center-of-mass subspace:
\begin{equation}
\mathbf{x}_{\text{centered}} = \mathbf{x} - \frac{1}{N}\sum_{i=1}^{N} \mathbf{x}_i,
\label{eq:com-constraint}
\end{equation}
reducing the effective coordinate dimensionality to:
\begin{equation}
d_{\text{subspace}} = (N - 1) \times 3.
\end{equation}
This constraint is necessary because the diffusion process operates in the translation-invariant subspace; without re-centering, accumulated numerical drift would shift the molecule away from the pocket centroid.

\subsection{Atom Type Handling}

During diffusion, atom types are treated as continuous vectors in a 9-dimensional space (corresponding to the atom vocabulary \{C, N, O, F, P, S, Cl, Br, I\}). Gaussian noise is added to the one-hot encoding during the forward process, and the reverse process predicts continuous logits.

\paragraph{Discretization at $t = 0$.}
At the final denoising step, continuous predictions are discretized via:
\begin{equation}
P(h = c \mid \mathbf{z}_0) \propto \Phi\!\left(\frac{c + 0.5 - z_h}{\sigma_0}\right) - \Phi\!\left(\frac{c - 0.5 - z_h}{\sigma_0}\right),
\label{eq:atom-discretization}
\end{equation}
where:
\begin{itemize}[nosep,leftmargin=1.5em]
    \item $\Phi(\cdot)$ -- standard normal CDF
    \item $z_h$ -- predicted continuous atom-type logit
    \item $\sigma_0$ -- residual noise level at $t = 0$
    \item $c \in \{0, 1, \ldots, 8\}$ -- index over the 9 atom types
\end{itemize}

\subsection{Training Loss}

The full training objective combines four terms:
\begin{equation}
\mathcal{L} = \underbrace{-\log p(\mathbf{x})}_{\text{subspace}} + \underbrace{\left(1 - e^{-(\gamma_s - \gamma_t)}\right)\,\|\hat{\boldsymbol{\epsilon}} - \boldsymbol{\epsilon}\|^2}_{\text{SNR-weighted MSE}} + \underbrace{\mathcal{L}_0}_{\text{reconstruction}} + \underbrace{D_{\text{KL}}\!\left[q(\mathbf{z}_T|\mathbf{x}) \,\|\, p(\mathbf{z}_T)\right]}_{\text{KL prior}},
\label{eq:training-loss}
\end{equation}
where:
\begin{itemize}[nosep,leftmargin=1.5em]
    \item $\gamma_t = \log(\sigma_t^2 / \alpha_t^2)$ -- negative log signal-to-noise ratio
    \item The SNR-weighting ensures each noise level contributes proportionally to its informational content
    \item $\mathcal{L}_0$ -- reconstruction loss evaluating the quality of the final denoised sample
    \item The KL prior term ensures the fully noised distribution matches $\mathcal{N}(\mathbf{0}, \mathbf{I})$
\end{itemize}

\subsection{Multi-Objective Scoring (Expanded)}
\label{appendix:multi-obj}

\paragraph{Adaptive Threshold Weight Update.}
Objective weights are adjusted after each generation batch based on satisfaction rates:
\begin{equation}
w_i^{(k+1)} = (1 - \alpha)\, w_i^{(k)} + \alpha \cdot \frac{1 - s_i}{\sum_{j} (1 - s_j)},
\label{eq:adaptive-weights}
\end{equation}
where:
\begin{itemize}[nosep,leftmargin=1.5em]
    \item $s_i \in [0,1]$ -- fraction of samples satisfying objective $i$'s threshold
    \item $\alpha = 0.1$ -- adaptation rate
    \item Weights are bounded to $[0.1, 0.6]$; SA maintains a minimum weight of 0.25
\end{itemize}
This inverse-satisfaction weighting up-weights objectives that are harder to satisfy, dynamically re-balancing the optimization landscape.

\paragraph{Individual Objective Losses.}
\begin{align}
\mathcal{L}_{\text{aff}} &= (\text{pKd}_{\text{target}} - \text{pKd})^2, \label{eq:loss-aff} \\
\mathcal{L}_{\text{dock}} &= (\text{Vina} - \text{Vina}_{\text{target}})^2, \label{eq:loss-dock} \\
\mathcal{L}_{\text{SA}} &= 10 \cdot \left(\exp(\text{SA} - \text{SA}_{\max}) - 1\right) + (\text{SA} - \text{SA}_{\text{target}})^2, \label{eq:loss-sa}
\end{align}
where SA$_{\max} = 3.5$ is the maximum acceptable SA score, pKd$_{\text{target}} = 7.0$, and Vina$_{\text{target}} = -11.0$~kcal/mol. The SA loss includes an exponential penalty term (multiplier $= 10.0$) that sharply penalizes molecules exceeding the SA threshold.

\paragraph{Combined Loss.}
\begin{equation}
\mathcal{L}_{\text{combined}} = \sum_i w_i\, \mathcal{L}_i.
\end{equation}

\subsection{Proxy Affinity Score (HNN-Denovo)}
\label{sec:proxy}

Since HNN-Denovo operates on SMILES rather than 3D coordinates, gradient guidance uses a differentiable proxy affinity score computed from atomic positions:
\begin{align}
\text{score}_{\text{distance}} &= \frac{10.0}{1 + d_{\text{centroid}} / 10.0}, \label{eq:proxy-distance} \\
\text{score}_{\text{contact}} &= 0.3 \sum_{i} \sigma\!\left((\tau - d_i^{\min}) \cdot \kappa \right), \label{eq:proxy-contact} \\
\text{score}_{\text{proximity}} &= \frac{5.0}{1 + d_{\min}^{\text{pocket}} / 4.0}, \label{eq:proxy-proximity}
\end{align}
where:
\begin{itemize}[nosep,leftmargin=1.5em]
    \item $d_{\text{centroid}}$ -- distance from ligand geometric center to the pocket centroid
    \item $d_i^{\min}$ -- minimum distance from ligand atom $i$ to any pocket atom
    \item $d_{\min}^{\text{pocket}}$ -- minimum distance from any ligand atom to any pocket atom
    \item $\sigma(\cdot)$ -- sigmoid function
    \item $\tau = 4.5$~\AA{} -- contact distance threshold
    \item $\kappa = 2.0$ -- sigmoid sharpness parameter
\end{itemize}
The combined proxy score is:
\begin{equation}
\text{proxy\_affinity} = \text{score}_{\text{distance}} + \text{score}_{\text{contact}} + \text{score}_{\text{proximity}}.
\end{equation}
All components are differentiable with respect to ligand coordinates, enabling direct backpropagation without finite-difference approximation. This proxy achieves a Pearson correlation of $r = 0.224$ against Vina ground truth (Section~\ref{sec:hnn}). Crucially, this reflects the quality of the proxy \emph{as a surrogate for Vina}, not the accuracy of the HNN-Denovo model itself: HNN-Denovo achieves PCC~$= 0.72$ predicting pKd from SMILES and protein sequences, but that capability cannot be directly leveraged for 3D gradient guidance without an intermediate coordinate-based approximation.

\end{document}